\documentclass[twocolumn]{aastex631}
\usepackage{hyperref}
\usepackage{breakurl}
\shorttitle{Relativistic Shock-Clump Interaction}
\shortauthors{Tomita et al.}
\graphicspath{{./}{figures/}}

\begin{document}

\title{Interaction of a Relativistic Magnetized Collisionless Shock with a Dense Clump}

\author[0000-0001-7952-2474]{Sara Tomita}
\affiliation{Frontier Research Institute for Interdisciplinary Sciences, Tohoku University,\\
Sendai, 980-8578, Japan}
\affiliation{Astronomical Institute, Graduate School of Science, Tohoku University,\\
Sendai, 980-8578, Japan}

\author[0000-0002-2387-0151]{Yutaka Ohira}
\affiliation{Department of Earth and Planetary Science, The University of Tokyo, 7-3-1 Hongo, Bunkyo-ku, Tokyo 113-0033, Japan}


\author[0000-0003-2579-7266]{Shigeo S. Kimura}
\affiliation{Frontier Research Institute for Interdisciplinary Sciences, Tohoku University,\\
Sendai, 980-8578, Japan}
\affiliation{Astronomical Institute, Graduate School of Science, Tohoku University,\\
Sendai, 980-8578, Japan}

\author[0000-0001-8105-8113]{Kengo Tomida}
\affiliation{Astronomical Institute, Graduate School of Science, Tohoku University,\\
Sendai, 980-8578, Japan}

\author[0000-0002-7114-6010]{Kenji Toma}
\affiliation{Frontier Research Institute for Interdisciplinary Sciences, Tohoku University,\\
Sendai, 980-8578, Japan}
\affiliation{Astronomical Institute, Graduate School of Science, Tohoku University,\\
Sendai, 980-8578, Japan}



\begin{abstract}
The interactions between a relativistic magnetized collisionless shock and dense clumps have been expected to play a crucial role on the magnetic field amplification and cosmic-ray acceleration. We investigate this process by two-dimensional Particle-In-Cell (PIC) simulations for the first time, where the clump size is much larger than the gyroradius of downstream particles. 
We also perform relativistic magnetohydrodynamic (MHD) simulations for the same condition to see the kinetic effects.
We find that particles escape from the shocked clump along magnetic field lines in the PIC simulations, so that the vorticity is lower than that in the MHD simulations. Moreover, in both the PIC and MHD simulations, the shocked clump quickly decelerates because of relativistic effects. Owing to the escape and the deceleration, the shocked clump cannot amplify the downstream magnetic field in relativistic collisionless shocks.
This large-scale PIC simulation opens a new window to understand large-scale behaviors in collisionless plasma systems.
\end{abstract}



\section{Introduction} 
Relativistic shocks are formed in high-energy astrophysical phenomena such as gamma-ray bursts, relativistic jets from active galactic nuclei, and pulsar wind nebulae.
In high-temperature astrophysical plasmas, shocks are often collisionless in the sense that Coulomb interactions do not play an important role in the shock dissipation. 
Instead, shocks are mediated by collective motion of charged particles which generates fluctuation of electromagnetic fields, disturbing particle orbits. 
In addition, some particles are accelerated to very large energies compared with the downstream temperature in the collisionless shock \citep{spitkovsky08a,sironi13}. 
Therefore, the collisionless shock is believed to accelerate cosmic rays and emit powerful non-thermal radiations, and is expected to have a crucial role in various high-energy astrophysical phenomena \citep{kotera11,tanaka11,fang12,murase12,kakuwa15,kimura18,zhang18,heinze20}.

The magnetic field is expected to be amplified in the collisionless shock to enhance the acceleration rate and radiation efficiency of accelerated electrons  \citep[e.g.][]{uchiyama07,magic19a,magic19b,lhaaso21,breuhaus22}.
For example, afterglow observations of Gamma-ray bursts suggest that the downstream magnetic field is much larger than the shock compressed value \citep[and reference therein]{santana14,tomita19,abdalla19}.
In collisionless shocks, kinetic plasma instabilities are induced, amplifying the magnetic field \citep{weibel59,lucek00,ohira09}.
In relativistic collisionless shocks, the Weibel instability rapidly generate the strong magnetic field fluctuation \citep{weibel59,kato05}, whose characteristic wavelength is the plasma skin-depth and much smaller than the astrophysical scale. 
Although the Weibel instability is required for particle acceleration in the relativistic shock \citep{niemiec06}, such a small-scale magnetic field rapidly decays near the shock front \citep{spitkovsky08b}. 
Therefore, in addition to the Weibel instability, other amplification mechanism of the magnetic field is required \citep{keshet09,tomita16,tomita19}.

Shock waves generally propagate into inhomogeneous media. For example,
the interaction of shock waves with a dense clump has been investigated
by hydrodynamical simulations and laboratory experiments for a long time \citep{stone92,atzeni04,inoue09,nishihara10,sano13,hennebelle19,sano21,zhou21,perkins17}.
Past studies show that in magnetized plasmas with a sufficiently large conductivity, the shock-clump interactions generate turbulence that amplifies magnetic fields. 
The turbulent dynamo in the shock downstream region is expected to play a crucial role on the magnetic field amplification, cosmic-ray acceleration, and enhancement of non-thermal radiation in many astrophysical objects \citep{sironi07,giacalone07,inoue11,mizuno11,fraschetti13,mizuno14}. 
Recently, physical processes of collisionless shocks is quite actively investigated by Particle-In-Cell (PIC) simulations \citep{spitkovsky08a,spitkovsky08b,niemiec12,matsumoto13,iwamoto19} but the upstream medium is usually assumed to be uniform.  Although there are a few exceptional studies \citep{sironi12,tomita16,tomita19}, the upstream inhomogeneity was assumed to be a simple structure like a plane wave.

Since the mean free path of Coulomb collision is longer than the system size in collisionless shock, 
particles can easily escape from the dense clump by diffusion and free streaming motion along a magnetic field line. 
After the dense clump passes through the collisionless shock front, the thermal velocity in the clump becomes high, 
so that particles in the clump would escape in the sound crossing time \citep{tomita16}.  
However, this escape process has not been taken into account in early studies on the shock-clump interaction by using magnetohydrodynamic (MHD) simulations \citep{shin08,inoue11,mizuno11,sano13,mizuno14}. 
Therefore, it is an open problem whether or not the shock-clump interaction in collisionless plasmas can amplify the downstream magnetic field by the turbulent dynamo.
In this Letter, we present the first {\it ab initio} PIC simulation of the shock-clump interaction in collisionless plasmas, and perform MHD simulations for the same condition to see the kinetic effects. 

%
\section{Simulation Setup}
We perform fully kinetic two-dimensional simulations of relativistic shocks propagating into inhomogeneous media using electromagnetic PIC code \citep{matsumoto13,ikeya15,matsumoto15}. 
The simulation frame and box are set in the downstream rest frame and the $x-y$ plane. 
In this PIC simulation, particles are continuously injected with a drift velocity in the $x$-direction from one side of the simulation boundary, and reflected at the opposite side. 
We apply the moving injection boundary to reduce computational costs and the numerical heating by numerical Cherenkov instability \citep{godfrey74}. 
The periodic boundary condition is assumed in the $y$-direction. 
The initial upstream density distribution in the simulation frame is given by
\begin{eqnarray}
n(x,y) &=& \left \{ \begin{array}{ll}
 n_0 & (r >  2R_{\rm cl} )\nonumber \\
n_0 + (n_{\rm cl}-n_0)\left\{1 +  \cos\left(\frac{\pi r}{2R_{\rm cl}}\right)\right\} &(r \leqq  2R_{\rm cl} ) \nonumber \\
\end{array} \right., \nonumber \\
r &\equiv&\sqrt{\frac{(x-x_{\rm c})^2}{\Gamma^2} + (y-y_{\rm c})^2 }, \nonumber
\end{eqnarray}
where $n_0$ and $\Gamma$ are the number density in the uniform region and the bulk Lorentz factor of the upstream plasma, 
and $x_{\rm c}$ and $y_{\rm c}$ are $x$ and $y$ coordinates at the clump center. 
$n_{\rm cl}$ is the number density at the half width at half maximum of the clump, $R_{\rm cl}$. 
Since we consider a spherical clump structure with the radius of $R_{\rm cl}$ in the upstream rest frame, its structure is compressed along the $x$-direction due to the Lorentz contraction in the simulation frame.
The density structure is not stable because the temperature is uniform.
However, the thermal velocity in the upstream region is much smaller than the speed of light, so that the density structure in our PIC simulation does not change significantly until the clump interacts with the shock front.

According to previous simulations, electrons are heated to near equipartition with ions in the downstream region for relativistic shocks \citep{spitkovsky08b,kumar15}, so that electron-ion plasmas can be approximately treated by electron-positron plasmas. 
Hence, we consider electron-positron plasmas to reduce computational costs in this study. 
The density, $n$, represents the total density of electrons and positrons. 
The uniform magnetic field parallel to the $y$-direction is imposed in the upstream region, ${\vec B} = B_0 {\vec e}_y$.
The magnetic field strength, $B_0$, is characterized by the plasma magnetization parameter, $\sigma_{\rm e}=B_0^2/(8 \pi \Gamma n_0 m_{\rm e} c^2) $, 
where $m_{\rm e}$ and $c$ are the electron mass and speed of light. 
The simulation box size is $L_x \times L_y = 3120c/\omega_{\rm pe} \times 1200c/\omega_{\rm pe}$ at the end of simulation, 
$t_{\rm end}=6300\omega_{\rm pe}^{-1}$, where $\omega_{\rm pe}= (4\pi n_{\rm 0} e^2/\Gamma m_{\rm e})^{1/2}$ is the plasma frequency of electron-positron plasma 
and $e$ is the elementary charge. 
The cell size and time step of simulations are $\Delta x = \Delta y = 0.1c/ \omega_{\rm pe}$ and $\Delta t = 0.1\omega_{\rm pe}^{-1}$, respectively\footnote{In order to suppress the numerical Cherenkov instability in PIC simulations\citep{godfrey74}, Maxwell ’s equations are solved by an implicit method with the CFL number of 1.0 \citep{ikeya15}. Thanks to the implicit method, there is no side effects.}. 
The number of simulation particles is 60 in each cell of the uniform region for electrons and positrons.

We perform several simulations with a different density of the clump, $n_{\rm cl}$, whose values are 1.5, 2.0, 4.0, 6.0, 8.0, 10.0, and 20.0, respectively. 
The other parameters and the thermal velocity of the upstream plasma are fixed to be $R_{\rm cl}=300c/\omega_{\rm pe}, \Gamma=10, \sigma_{\rm e}=10^{-3}$, and $v_{\rm th}=0.18c$ in this work.
Then, the ratio of the clump size to the gyroradius \footnote{Coincidentally, in this work, the downstream gyroradius is the same order of magnitude as the upstream gyroradius, $r_{\rm g} = 1.39 r_{\rm g,u}$, where $r_{\rm g,u}$ is the upstream gyroradius in the upstream rest frame.} of the downstream thermal electrons and positron is $R_{\rm cl}/r_{\rm g}=53.7\gg1$. 
This indicates that the downstream plasma is expected to be well magnetized in the clump scale.  

%
%
To clarify the effect of particle escape on the turbulent dynamo, we also perform MHD simulations for the shock-clump interaction with the open source code, Athena++ \citep{white16,stone20}, where we use the relativistic HLLE Riemann solver, 2nd-order piecewise linear reconstruction,
and 2nd-order Runge-Kutta time integrator.
All the physical parameters are the same as those in the PIC simulations. 
The cell size and time step of MHD simulations are $\Delta x = 0.1\Delta y = R_{\rm cl}/3000$, and $\Delta t = 0.1\Delta x/c$, respectively. 
The density structure in the MHD simulation is stable because the pressure is uniform, that is, the structure is in the dynamic equilibrium state.

\section{Results}


\begin{figure}
{\centering  
{\includegraphics[width=5.5cm, angle=-90]{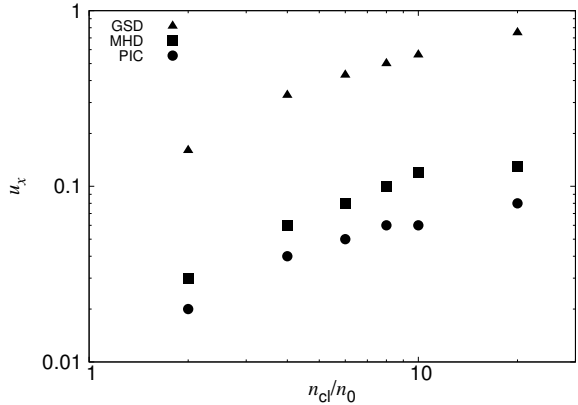}}
\par}
\caption{The $x$ component of four velocity at the high-density clump, $u_x$, at $t=2t_{\rm esc}$ as a function of the density of clump, $n_{\rm cl}/n_0$. The triangle, square, and circle points show results for GSD approximation, the MHD simulation, and the PIC simulation, respectively. }  \label{f5}
\end{figure}
\begin{figure*}[t]
{\centering
{\includegraphics[width=11cm]{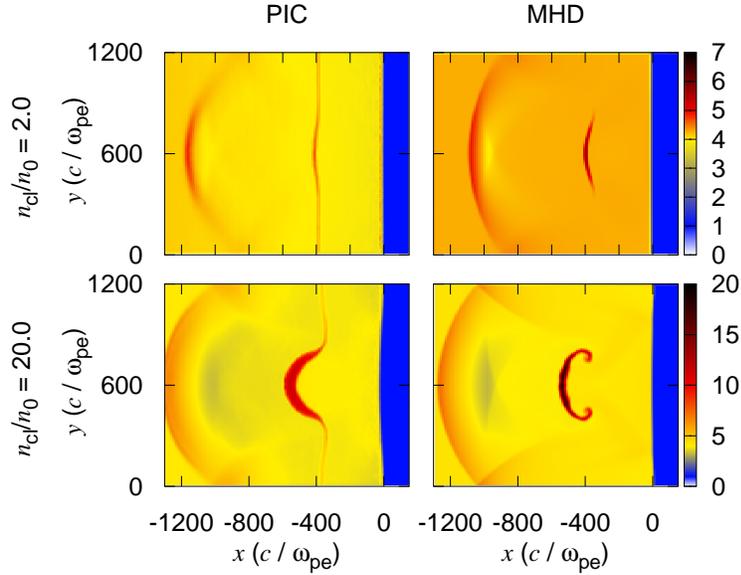}}
\par}
\caption{Density distribution, $n/n_0$, in the shock downstream region at $t=2t_{\rm esc}$ for $n_{\rm cl}/n_0=2$ (top two panels) and $n_{\rm cl}/n_0=20$ (bottom two panels). The time, $t$, represents the elapsed time since the clump interacts with the shock front. The left and right columns show results for the PIC and MHD simulations, respectively. Movie: \url{https://docs.google.com/presentation/d/1cseAc5ehoR-9V917zVC_lzvVw9Pi3a8Q/edit?usp=sharing&ouid=112346314884784410578&rtpof=true&sd=true}} \label{f1}
\end{figure*}
\begin{figure*}[t]
{\centering
{\includegraphics[width=11cm]{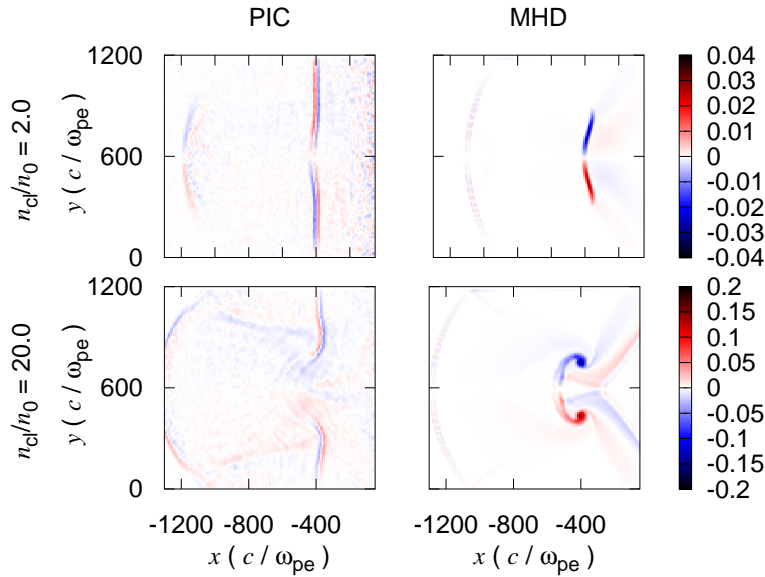}}
\par}
\caption{Same as Fig.~\ref{f1}, but for the $z$ component of vorticity, $(\vec{\nabla} \times {\vec u})_z$.}  \label{f2}
\end{figure*}
\begin{figure*}[t]
\begin{minipage}[b]{0.45\linewidth}
\includegraphics[width=9.5cm]{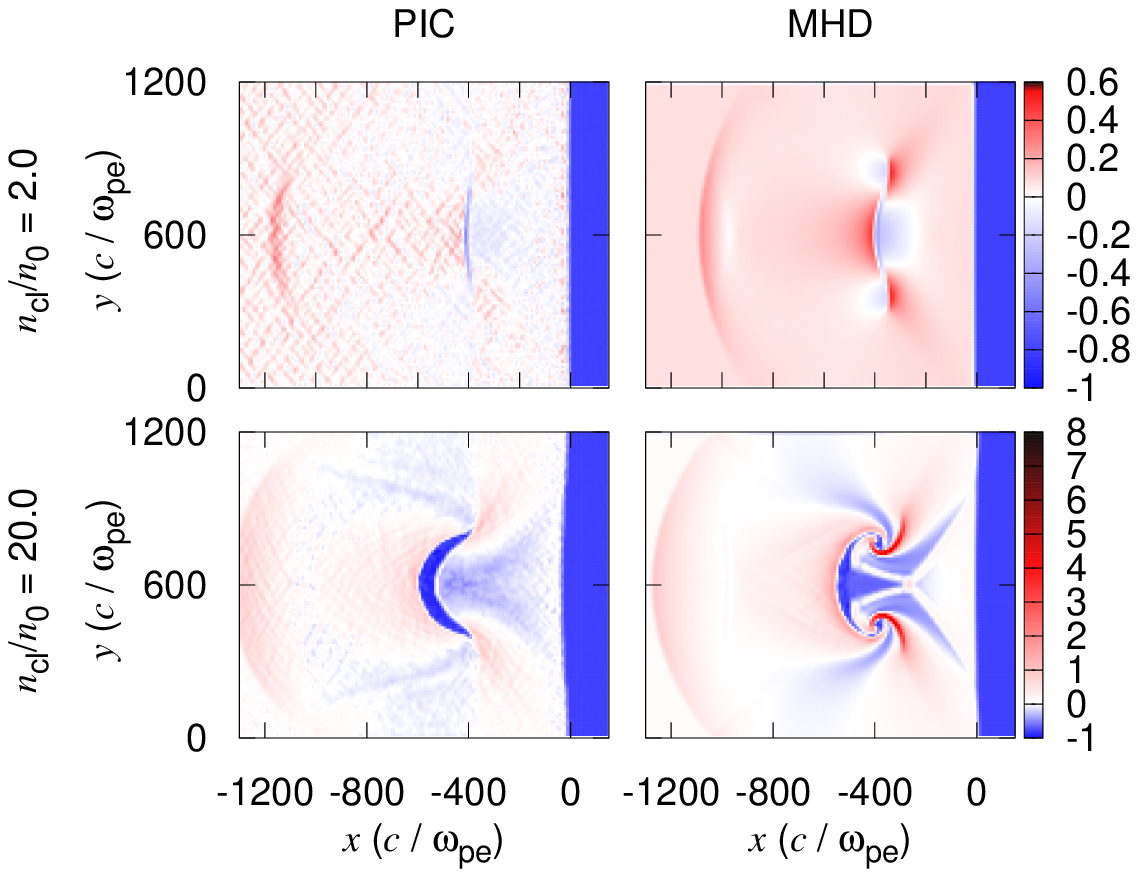}
\end{minipage}
\begin{minipage}[b]{0.45\linewidth}
\hspace*{+0.8cm} 
\includegraphics[width=9.5cm]{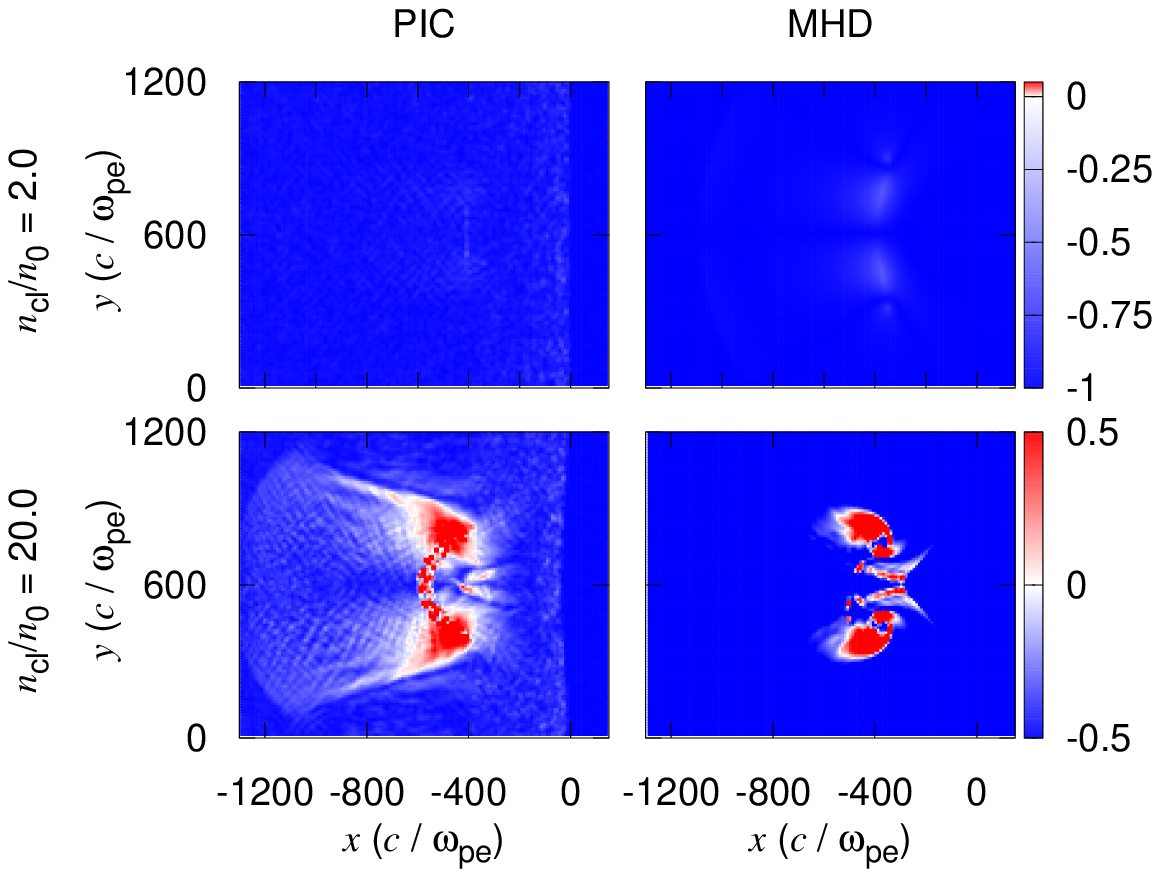}
\end{minipage}
\caption{Same as Fig.~\ref{f1}, but for the magnetic field strength, $(B-4B_0)/4B_0$ (left column) and the ratio of $B_x$ to $B_y$, $|B_x/B_y|-1$ (right column).} \label{f3}
\end{figure*}
\begin{figure}[t]
{\centering
{\includegraphics[width=5.5cm, angle=-90]{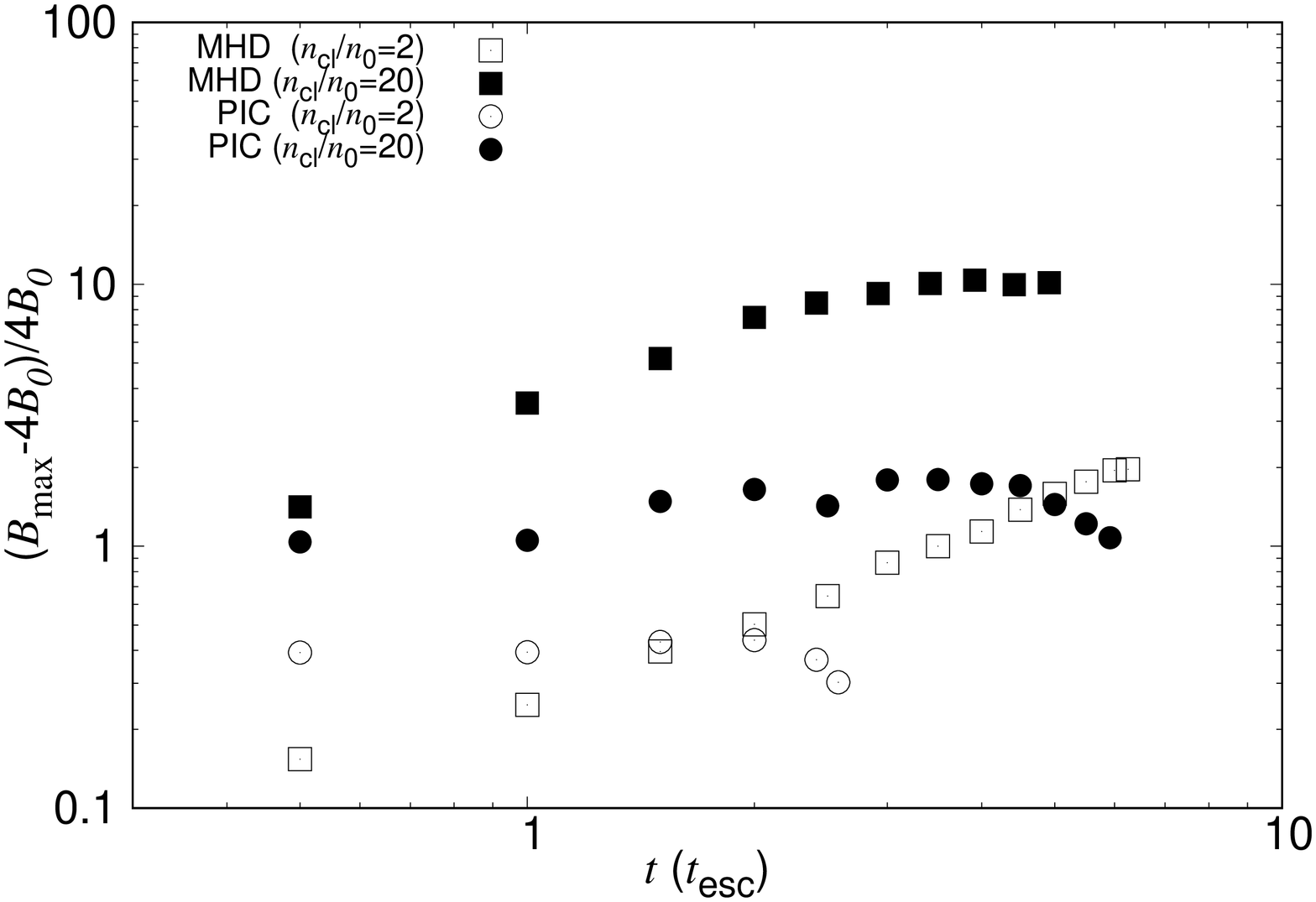}}
\par 
}
\caption{Time evolution of the maximum magnetic-field amplification factor in the shock downstream region, $(B_{\rm max}-4B_0)/4B_0$. The time, $t$, represents the elapsed time since the clump interacts with the shock front. The open and filled points show results for $n_{\rm cl}/n_0=2$ and $20$, respectively. The square and circle points show results for the MHD and PIC simulations, respectively.}  \label{f4}
\end{figure}

Fig.~\ref{f5} shows the $x$ component of four velocity at the high-density clump, $u_x$, at $t=2t_{\rm esc}$ as a function of the density of the clump, $n_{\rm cl}/n_0$.  
The time, $t$, represents the elapsed time since the clump interacts with the shock front. 
$t_{\rm esc}=R_{{\rm cl}}/0.5c=600\omega_{\rm pe}^{-1}$ is the escape time along the magnetic field line ($y$-direction), where the mean particle velocity along the magnetic field line is assumed to be $0.5c$.
In addition to the MHD and PIC results, theoretical expectations based on the geometrical shock dynamics (GSD) \citep{sironi07} are plotted in Fig.~\ref{f5}.
The GSD approximation assumed that the downstream flow is determined by the forward-going Riemann characteristics in isentropic and homogeneous far downstream region.
Our MHD simulations show that the clump velocities in the downstream region are much lower than those expected from the GSD approximation. 
\citet{sironi07} estimated the downstream velocity disturbance due to the clump-shock interaction based on the GSD approximation without taking into account the deceleration of shocked clump. 
However, we observed the significant deceleration of the shocked clump in the downstream region. 

The reason for this rapid deceleration is as follows.
At first, the clump decelerates by the shock interaction in the shock transition time. After that, the shocked clump decelerates further as propagating to the uniform downstream region. Since the velocity of the shocked clump in the downstream rest frame is non-relativistic after the shock crossing, the temperature in the shocked clump is highly relativistic. Then, the momentum of the shocked clump in the downstream rest frame is approximately
\begin{equation}
P_{\rm cl}=\frac{4}{3}\Gamma M_{\rm cl}\gamma^2 v_{\rm cl,d}, \nonumber
\end{equation}
where $v_{\rm cl,d}$ and $\gamma$ are the three-velocity of the shocked clump and the corresponding Lorentz factor, respectively. The momentum flux that the clump sweeps is  
\begin{equation}
F=(\epsilon + p)\gamma^2 \frac{v_{\rm cl,d}^2}{c^2} R_{\rm cl}^2, \nonumber
\end{equation}
where $\epsilon =4\Gamma^2 n^{\prime}m_{\rm p}c^2$ and $p$ are the fluid energy density and pressure in the downstream uniform region, $p\approx\epsilon/3$ for relativistically hot plasmas. 
$R_{\rm cl}^2$ is the cross section of the clump.
$R_{\rm cl}, n^{\prime}$ and $m_{\rm p}$ are the comoving clump size, comoving number density in the upstream uniform region and the proton mass, respectively. 
Then, the deceleration time is given by
\begin{equation}
t_{\rm dec} = \frac{P_{\rm cl}}{F}=\frac{M_{\rm cl}}{\Gamma n^{\prime} m_{\rm p}v_{\rm cl,d}R_{\rm cl}^2}. \label{eq3}
\end{equation}
In the above argument, we consider a clump in the three-dimensional space.
In the two-dimensional space like our simulations, the cross section changes to $R_{\rm cl}$,
but the relation of $t_{\rm dec} \propto \Gamma^{-1}$, does not change.
We found that, in relativistic shocks, the deceleration time of the clump is $\Gamma$ times shorter than that in non-relativistic shocks. 
The fluid energy density $\epsilon$ in the downstream uniform region is a crucial factor for the rapid deceleration.
The factor of $\Gamma^2$ originates from the Lorentz contraction concerning the number density and the random kinetic energy of each particle in the downstream region.
Previous MHD simulations were performed in the mildly relativistic flow with upstream bulk Lorentz factor of about 2. 
Thus, this effect is neglected in previous studies.

In the PIC simulations, the clump velocity is decreased further.
This is because the deceleration time in the PIC simulations is shorter than that in the MHD simulations since the cross section of the clump becomes large due to the streaming of particles.


Fig.~\ref{f1} shows density in the shock downstream region at $t=2t_{\rm esc}$. 
The shock front is located at $x=0$ and the left region ($x<0$) is the shock downstream region. 
In the MHD simulation with $n_{\rm cl}/n_0=2$ (top right), the structure of the high-density clump is not significantly deformed by the downstream shear flow even though the density of the clump is not so low. 
By inserting $M_{\rm cl}\approx n_{\rm cl}^{\prime}R_{\rm cl}^3$ to Eq.~\ref{eq3}, the deceleration time is represented by 
\begin{equation}
t_{\rm dec} = \frac{n_{\rm cl}^{\prime}}{\Gamma n^{\prime}}\frac{R_{\rm cl}}{v_{\rm cl,d}} = \frac{n_{\rm cl}^{\prime}}{\Gamma n^{\prime}} t_{\rm eddy},
\end{equation}
where $t_{\rm eddy}=R_{\rm cl}/v_{\rm cl,d}$, and $n_{\rm cl}^{\prime}$ is the comoving number density of the upstream clump. For $n_{\rm cl}/n_0=2$, the deceleration time is shorter than the eddy turnover time ($t_{\rm dec}<t_{\rm eddy}$), so that the clump decelerates in the downstream region before the clump is deformed by the downstream shear flow. On the other hand, for $n_{\rm cl}/n_0=20$, the deceleration time is longer than the eddy turnover time ($t_{\rm dec}>t_{\rm eddy}$), so that the clump structure is strongly deformed.
In the PIC simulations (left two panels), particles in the clump escape along the magnetic field line because particles get a large velocity dispersion after the shock heating. 
The density becomes almost uniform along the magnetic field line in the downstream region for $n_{\rm cl}/n_0=2$ (top left). 
In the PIC simulation with $n_{\rm cl}/n_0=20$ (bottom left), although the density is not completely uniform along the magnetic field line, the deformation of the clump structure is smaller than that in the MHD simulation.
The high-density structures at $x=-1200c/\omega_{\rm pe}$ for all the cases are sound waves excited by the shock-clump interaction, which develop to a shock-like structure due to the nonlinear steepening. 

Fig.~\ref{f2} shows the $z$ component of vorticity, $(\vec{\nabla} \times {\vec u})_z$, at $t=2t_{\rm esc}$. 
The vorticity in the PIC simulations (left two panels) is significantly lower than one in the MHD simulations. 
As mentioned in Fig.~\ref{f5}, the shocked clump decelerates more rapidly in the PIC simulations because the particle streaming along the magnetic field line makes the cross section of the clump large. 
Since the vorticity is estimated by the clump velocity over the clump size, the vorticity is low for the PIC simulations.
%

The magnetic field strength, $|{\vec B}|$, at $t=2t_{\rm esc}$ is given in the left column in Fig.~\ref{f3}.
As expected from the vorticity shown in Fig.~\ref{f2}, the magnetic field is strongly amplified by the vortex motion in the the MHD simulation with $n_{\rm cl}/n_0=20$ but not in the PIC simulations. 
The time evolution of the maximum strength of the downstream magnetic fields, $B_{\rm max}$, is shown in Fig.~\ref{f4}. 
Except for the PIC simulation with $n_{\rm cl}/n_0=2$, the maximum field strength is at least three times larger than the shock compressed value of $4B_0$, i.e., $B_{\rm max}\gtrsim12B_0$.
Moreover, the $x$ component of the amplified field is comparable to or larger than $4B_0$ (see the right column in Fig.~\ref{f3}), which is generated by stretching of the magnetic field line. 
The shock compression and downstream sound waves amplifies only the $y$ component of the magnetic field. 
Therefore, the growth of the magnetic field strength is due to the vortex motion.
For $n_{\rm cl}/n_0=2$, although the magnetic field continues to be amplified in the MHD simulations for a long time, it saturates at $t\sim 2t_{\rm esc}=1200\omega_{\rm pe}^{-1}$ in the PIC simulation, where $B_{\rm max}\sim 6B_0$.
Since the saturation time is about $2 t_{\rm esc}$ in the PIC simulation with $n_{\rm cl}/n_0=2$, the saturation of the magnetic field amplification originates from the particle escape. 
For a higher-density clump ($n_{\rm cl}/n_0=20$), the magnetic field is amplified to the equipartition level with the upstream kinetic energy in the MHD simulation, that is, $B_{\rm max}\sim B_0\sqrt{\sigma_{\rm e}}^{-1}\sim31.6 B_0$.
On the other hand, in the PIC simulation, the growth rate is lower than that in the MHD simulation. Moreover, the magnetic field saturates before reaching the equipartition level.
Therefore, our PIC simulations show that the downstream magnetic field is not efficiently amplified by the shock-clump interaction for relativistic collisionless shocks even 
though the clump density is much higher than the mean density, $n_{\rm cl}/n_0 =20$.

To summarize the results, for the PIC simulations, the particle streaming motion causes a large velocity gradient scale, $l$, and a small velocity perturbation, $\delta u$. This leads to a very low vorticity around the shocked clump, $\delta u/l$, compared to that in the MHD simulations. This occurs even though the clump size is much larger than the gyroradius of downstream particles. 
Therefore, the magnetic field amplification is slower or saturates at a lower level in the PIC simulations. 
%
%
%
\section{Discussion}
%
We have considered a magnetized plasma with $\sigma_{\rm e}=10^{-3}$ in this work.
For more weakly magnetized plasmas (lower $\sigma_{\rm e}$), some kinetic plasma instabilities would be excited by the escape from the dense clump, which potentially generate magnetic field fluctuations \citep{tomita16,tomita19}.
Such kinetic-scale magnetic field fluctuations could disturb the free-streaming motion along the magnetic field line.
This could suppresses the streaming escape along the magnetic field line.
Even in collisionless systems, as long as the escape timescale due to diffusion or free streaming is longer than the eddy turnover time,
the shock-clump interaction would drive turbulent dynamo.
On the other hand, the particle diffusion perpendicular to the magnetic field would occur in a realistic three-dimensional system \citep{jokipii93,giacalone94},
so that it might be difficult to drive the turbulent dynamo by the shock-clump interaction.
We will address the shock-clump interaction for more weakly magnetized collisionelss shocks by means of three-dimensional PIC simulations in future work.

In our PIC simulations, the ratio of the clump size to the gyroradius of downstream thermal electrons and positrons is $R_{\rm cl}/r_{\rm g}=53.7$. 
In reality, the ratio of $R_{\rm cl}/r_{\rm g}$ is widely distributed in laboratory, space and astrophysical plasmas. 
Laboratory plasmas could have a similar value of $R_{\rm cl}/r_{\rm g}=\mathcal{O}(10)-\mathcal{O}(100)$, whereas it must be much larger than $\mathcal{O}(100)$ 
in astrophysical plasmas. 
The effect of escape and diffusion would depend on the ratio of $R_{\rm cl}/r_{\rm g}$.
If the diffusion length scale is much smaller than the size of the clump, 
the particle streaming along the magnetic field does not occur, but the rapid deceleration due to the relativistic effect shown in our relativistic MHD simulation occurs. 
The diffusion length scale cannot be easily estimated even in the simple system that we considered.
We need to conduct a parameter study for $R_{\rm cl}/r_{\rm g}$.

The clump structure in the downstream rest frame is likely important for driving the turbulent dynamo.
Large-scale hybrid plasma simulations show that non-relativistic collisionless shocks generate strong density fluctuations in the upstream region. These fluctuations results in the magnetic field amplification in the downstream, which may be interpreted as the shock-clump interactions \citep{caprioli13,ohira16a,ohira16b}.
For relativistic shocks, a spherical structure in the upstream rest frame is a structure compressed to the shock-normal direction in the downstream rest frame because of the the Lorentz contraction. 
Hence, the interaction time between the shock front and the dense clump is shorter, so that the velocity disturbance is smaller than that for nonrelativistic shocks. 
Furthermore, the compressed structure makes the deceleration time of the shocked clump short. 
Although the results are not shown here, we have confirmed by the PIC simulations that the magnetic field is amplified more efficiently by the interaction between a collisionless shock and a spherical clump in the shock downstream frame.
In addition, the particle streaming from a dense clump would be less important in nonrelativistic shocks because the thermal velocity in the dense clump can be much slower than the shock velocity for nonrelativistic shocks, but all velocity scales are always on the order of the speed of light for relativistic shocks.
Therefore, relativistic shocks are not as suitable for the turbulent dynamo caused by the shock-clump interaction as nonrelativistic shocks.

The rapid deceleration of the shocked clump has been observed in both MHD and PIC simulations. 
The most kinetic energy of the clump is quickly converted to the energy of sound waves, which eventually evolve to weak shocks. 
Since shocks interact with multiple clumps in reality, the sound waves or weak shocks interact with each other, resulting in a strong turbulence in the downstream region \citep[e.g.][]{inoue11}. 
Although such a nonlinear evolution has not been investigated in this work, 
in the future, we will perform a larger PIC simulations to understand the turbulent dynamo in the collisionless shock propagating to a more realistic nonuniform medium. 
In addition to the turbulent dynamo, particle acceleration by downstream turbulence could be observed in that simulation \citep{ohira13,pohl15,kimura16,zhdankin18,comisso19,kimura19,yokoyama20}. 


\section{Summary} 
In this work, we have performed the first PIC simulations for the shock-clump interaction in magnetized collisionless plasmas, 
finding that the downstream turbulence is significantly suppressed in collisionless shocks compared with results of the MHD simulations 
because particles escape from the dense clump region along the magnetic field line in the shock downstream region.
As a result, the magnetic field is not strongly amplified by the turbulent dynamo even though the upstream kinetic energy is much higher than that of the downstream magnetic field. 
Our simulation demonstrated that the particle streaming or diffusion have a significant impact on dynamics in the lengthscale larger than gyroradius. These processes cannot be handled by the MHD simulation. 
In addition, we have found that the dense clump quickly decelerates in the downstream region of relativistic shocks in the MHD and PIC simulations 
because the clump structure is compressed in the shock-normal direction by the Lorentz contraction. 
Thus, relativistic collisionless shocks are unsuitable for the downstream turbulent dynamo.

In many cases, we can only observe the emission from the shocked region. 
The spectra of the downstream magnetic field turbulence and accelerated particles are expected to depend on the upstream magnetization parameter and inhomogeneity. 
Conducting a parameter survey by the large-scale PIC simulations, we will be able to obtain the dependence of the downstream spectra of the non-thermal particles and magnetic field strength on the upstream plasma condition. 
Combining the PIC simulation results and observations, we can reveal the upstream plasma condition in the future.
Therefore, large-scale kinetic simulations incorporating the inhomogeneity of ambient media open a new window to understand laboratory, space, and astrophysical plasmas.

\section{Acknowledgments} 
We thank A. Kuwata, R. Kuze, Y. Matsumoto, and M. Kobayashi for useful comments.
The software used in this work was developed by Y. Matsumoto. 
Numerical computations were carried out on Cray XC50 at Center for Computational Astrophysics, National Astronomical Observatory of Japan. 
This work is supported by JSPS KAKENHI Grant Numbers 19H01893 (YO), 19J00198 (SSK), 21H04487 (K. Tomida and YO), 18H01245 (K. Toma). 
YO is supported by Leading Initiative for Excellent Young Researchers, MEXT, Japan. 
SSK acknowledges the support by the Tohoku Initiative for Fostering Global Researchers for Interdisciplinary Sciences (TI-FRIS) of MEXT's Strategic Professional Development Program for Young Researchers.

\bibliography{ms220718}{}

\begin{thebibliography}{}
\expandafter\ifx\csname natexlab\endcsname\relax\def\natexlab#1{#1}\fi
\providecommand{\url}[1]{\href{#1}{#1}}
\providecommand{\dodoi}[1]{doi:~\href{http://doi.org/#1}{\nolinkurl{#1}}}
\providecommand{\doeprint}[1]{\href{http://ascl.net/#1}{\nolinkurl{http://ascl.net/#1}}}
\providecommand{\doarXiv}[1]{\href{https://arxiv.org/abs/#1}{\nolinkurl{https://arxiv.org/abs/#1}}}

\bibitem[{{Abdalla} {et~al.}(2019){Abdalla}, {Adam}, {Aharonian}, {Ait
  Benkhali}, {Ang{\"u}ner}, {Arakawa}, {Arcaro}, {Armand}, {Ashkar}, {Backes},
  {Barbosa Martins}, {Barnard}, {Becherini}, {Berge}, {Bernl{\"o}hr},
  {Bissaldi}, {Blackwell}, {B{\"o}ttcher}, {Boisson}, {Bolmont}, {Bonnefoy},
  {Bregeon}, {Breuhaus}, {Brun}, {Brun}, {Bryan}, {B{\"u}chele}, {Bulik},
  {Bylund}, {Capasso}, {Caroff}, {Carosi}, {Casanova}, {Cerruti}, {Chand},
  {Chandra}, {Chen}, {Colafrancesco}, {Cury{\l}o}, {Davids}, {Deil}, {Devin},
  {deWilt}, {Dirson}, {Djannati-Ata{\"\i}}, {Dmytriiev}, {Donath},
  {Doroshenko}, {Dyks}, {Egberts}, {Emery}, {Ernenwein}, {Eschbach}, {Feijen},
  {Fegan}, {Fiasson}, {Fontaine}, {Funk}, {F{\"u}{\ss}ling}, {Gabici},
  {Gallant}, {Gat{\'e}}, {Giavitto}, {Giunti}, {Glawion}, {Glicenstein},
  {Gottschall}, {Grondin}, {Hahn}, {Haupt}, {Heinzelmann}, {Henri}, {Hermann},
  {Hinton}, {Hofmann}, {Hoischen}, {Holch}, {Holler}, {Horns}, {Huber},
  {Iwasaki}, {Jamrozy}, {Jankowsky}, {Jankowsky}, {Jardin-Blicq},
  {Jung-Richardt}, {Kastendieck}, {Katarzy{\'n}ski}, {Katsuragawa}, {Katz},
  {Khangulyan}, {Kh{\'e}lifi}, {King}, {Klepser}, {Klu{\'z}niak}, {Komin},
  {Kosack}, {Kostunin}, {Kreter}, {Lamanna}, {Lemi{\`e}re}, {Lemoine-Goumard},
  {Lenain}, {Leser}, {Levy}, {Lohse}, {Lypova}, {Mackey}, {Majumdar},
  {Malyshev}, {Marandon}, {Marcowith}, {Mares}, {Mariaud}, {Mart{\'\i}-Devesa},
  {Marx}, {Maurin}, {Meintjes}, {Mitchell}, {Moderski}, {Mohamed}, {Mohrmann},
  {Moore}, {Moulin}, {Muller}, {Murach}, {Nakashima}, {de Naurois},
  {Ndiyavala}, {Niederwanger}, {Niemiec}, {Oakes}, {O'Brien}, {Odaka}, {Ohm},
  {de Ona Wilhelmi}, {Ostrowski}, {Oya}, {Panter}, {Parsons}, {Perennes},
  {Petrucci}, {Peyaud}, {Piel}, {Pita}, {Poireau}, {Priyana Noel}, {Prokhorov},
  {Prokoph}, {P{\"u}hlhofer}, {Punch}, {Quirrenbach}, {Raab}, {Rauth},
  {Reimer}, {Reimer}, {Remy}, {Renaud}, {Rieger}, {Rinchiuso}, {Romoli},
  {Rowell}, {Rudak}, {Ruiz-Velasco}, {Sahakian}, {Sailer}, {Saito}, {Sanchez},
  {Santangelo}, {Sasaki}, {Schlickeiser}, {Sch{\"u}ssler}, {Schulz}, {Schutte},
  {Schwanke}, {Schwemmer}, {Seglar-Arroyo}, {Senniappan}, {Seyffert}, {Shafi},
  {Shiningayamwe}, {Simoni}, {Sinha}, {Sol}, {Specovius}, {Spir-Jacob},
  {Stawarz}, {Steenkamp}, {Stegmann}, {Steppa}, {Takahashi}, {Tavernier},
  {Taylor}, {Terrier}, {Tiziani}, {Tluczykont}, {Trichard}, {Tsirou}, {Tsuji},
  {Tuffs}, {Uchiyama}, {van der Walt}, {van Eldik}, {van Rensburg}, {van
  Soelen}, {Vasileiadis}, {Veh}, {Venter}, {Vincent}, {Vink}, {V{\"o}lk},
  {Vuillaume}, {Wadiasingh}, {Wagner}, {White}, {Wierzcholska}, {Yang},
  {Yoneda}, {Zacharias}, {Zanin}, {Zdziarski}, {Zech}, {Ziegler}, {Zorn},
  {{\.Z}ywucka}, {de Palma}, {Axelsson}, \& {Roberts}}]{abdalla19}
{Abdalla}, H., {Adam}, R., {Aharonian}, F., {et~al.} 2019, \nat, 575, 464,
  \dodoi{10.1038/s41586-019-1743-9}

\bibitem[{{Atzeni} \& {Meyer-ter-Vehn}(2004)}]{atzeni04}
{Atzeni}, S., \& {Meyer-ter-Vehn}, J. 2004, {The Physics of Inertial
  Fusion:Beam Plasma Interaction, Hydrodynamics, Hot Dense Matter},
  \dodoi{10.1093/acprof:oso/9780198562641.001.0001}

\bibitem[{{Breuhaus} {et~al.}(2022){Breuhaus}, {Reville}, \&
  {Hinton}}]{breuhaus22}
{Breuhaus}, M., {Reville}, B., \& {Hinton}, J.~A. 2022, \aap, 660, A8,
  \dodoi{10.1051/0004-6361/202142097}

\bibitem[{{Caprioli} \& {Spitkovsky}(2013)}]{caprioli13}
{Caprioli}, D., \& {Spitkovsky}, A. 2013, \apjl, 765, L20,
  \dodoi{10.1088/2041-8205/765/1/L20}

\bibitem[{{Comisso} \& {Sironi}(2019)}]{comisso19}
{Comisso}, L., \& {Sironi}, L. 2019, \apj, 886, 122,
  \dodoi{10.3847/1538-4357/ab4c33}

\bibitem[{{Fang} {et~al.}(2012){Fang}, {Kotera}, \& {Olinto}}]{fang12}
{Fang}, K., {Kotera}, K., \& {Olinto}, A.~V. 2012, \apj, 750, 118,
  \dodoi{10.1088/0004-637X/750/2/118}

\bibitem[{{Fraschetti}(2013)}]{fraschetti13}
{Fraschetti}, F. 2013, \apj, 770, 84, \dodoi{10.1088/0004-637X/770/2/84}

\bibitem[{{Giacalone} \& {Jokipii}(1994)}]{giacalone94}
{Giacalone}, J., \& {Jokipii}, J.~R. 1994, \apjl, 430, L137,
  \dodoi{10.1086/187457}

\bibitem[{{Giacalone} \& {Jokipii}(2007)}]{giacalone07}
---. 2007, \apjl, 663, L41, \dodoi{10.1086/519994}

\bibitem[{{Godfrey}(1974)}]{godfrey74}
{Godfrey}, B.~B. 1974, Journal of Computational Physics, 15, 504,
  \dodoi{10.1016/0021-9991(74)90076-X}

\bibitem[{{Heinze} {et~al.}(2020){Heinze}, {Biehl}, {Fedynitch}, {Boncioli},
  {Rudolph}, \& {Winter}}]{heinze20}
{Heinze}, J., {Biehl}, D., {Fedynitch}, A., {et~al.} 2020, \mnras, 498, 5990,
  \dodoi{10.1093/mnras/staa2751}

\bibitem[{{Hennebelle} \& {Inutsuka}(2019)}]{hennebelle19}
{Hennebelle}, P., \& {Inutsuka}, S.-i. 2019, Frontiers in Astronomy and Space
  Sciences, 6, 5, \dodoi{10.3389/fspas.2019.00005}

\bibitem[{{Ikeya} \& {Matsumoto}(2015)}]{ikeya15}
{Ikeya}, N., \& {Matsumoto}, Y. 2015, \pasj, 67, 64,
  \dodoi{10.1093/pasj/psv052}

\bibitem[{{Inoue} {et~al.}(2011){Inoue}, {Asano}, \& {Ioka}}]{inoue11}
{Inoue}, T., {Asano}, K., \& {Ioka}, K. 2011, \apj, 734, 77,
  \dodoi{10.1088/0004-637X/734/2/77}

\bibitem[{{Inoue} {et~al.}(2009){Inoue}, {Yamazaki}, \& {Inutsuka}}]{inoue09}
{Inoue}, T., {Yamazaki}, R., \& {Inutsuka}, S.-i. 2009, \apj, 695, 825,
  \dodoi{10.1088/0004-637X/695/2/825}

\bibitem[{{Iwamoto} {et~al.}(2019){Iwamoto}, {Amano}, {Hoshino}, {Matsumoto},
  {Niemiec}, {Ligorini}, {Kobzar}, \& {Pohl}}]{iwamoto19}
{Iwamoto}, M., {Amano}, T., {Hoshino}, M., {et~al.} 2019, \apjl, 883, L35,
  \dodoi{10.3847/2041-8213/ab4265}

\bibitem[{{Jokipii} {et~al.}(1993){Jokipii}, {Kota}, \&
  {Giacalone}}]{jokipii93}
{Jokipii}, J.~R., {Kota}, J., \& {Giacalone}, J. 1993, \grl, 20, 1759,
  \dodoi{10.1029/93GL01973}

\bibitem[{{Kakuwa} {et~al.}(2015){Kakuwa}, {Toma}, {Asano}, {Kusunose}, \&
  {Takahara}}]{kakuwa15}
{Kakuwa}, J., {Toma}, K., {Asano}, K., {Kusunose}, M., \& {Takahara}, F. 2015,
  \mnras, 449, 551, \dodoi{10.1093/mnras/stv281}

\bibitem[{{Kato}(2005)}]{kato05}
{Kato}, T.~N. 2005, Physics of Plasmas, 12, 080705, \dodoi{10.1063/1.2017942}

\bibitem[{{Keshet} {et~al.}(2009){Keshet}, {Katz}, {Spitkovsky}, \&
  {Waxman}}]{keshet09}
{Keshet}, U., {Katz}, B., {Spitkovsky}, A., \& {Waxman}, E. 2009, \apjl, 693,
  L127, \dodoi{10.1088/0004-637X/693/2/L127}

\bibitem[{{Kimura} {et~al.}(2018){Kimura}, {Murase}, \& {Zhang}}]{kimura18}
{Kimura}, S.~S., {Murase}, K., \& {Zhang}, B.~T. 2018, \prd, 97, 023026,
  \dodoi{10.1103/PhysRevD.97.023026}

\bibitem[{{Kimura} {et~al.}(2016){Kimura}, {Toma}, {Suzuki}, \&
  {Inutsuka}}]{kimura16}
{Kimura}, S.~S., {Toma}, K., {Suzuki}, T.~K., \& {Inutsuka}, S.-i. 2016, \apj,
  822, 88, \dodoi{10.3847/0004-637X/822/2/88}

\bibitem[{{Kimura} {et~al.}(2019){Kimura}, {Tomida}, \& {Murase}}]{kimura19}
{Kimura}, S.~S., {Tomida}, K., \& {Murase}, K. 2019, \mnras, 485, 163,
  \dodoi{10.1093/mnras/stz329}

\bibitem[{{Kotera} \& {Olinto}(2011)}]{kotera11}
{Kotera}, K., \& {Olinto}, A.~V. 2011, \araa, 49, 119,
  \dodoi{10.1146/annurev-astro-081710-102620}

\bibitem[{{Kumar} {et~al.}(2015){Kumar}, {Eichler}, \& {Gedalin}}]{kumar15}
{Kumar}, R., {Eichler}, D., \& {Gedalin}, M. 2015, \apj, 806, 165,
  \dodoi{10.1088/0004-637X/806/2/165}

\bibitem[{{Lhaaso Collaboration} {et~al.}(2021){Lhaaso Collaboration}, {Cao},
  {Aharonian}, {An}, {Axikegu}, {Bai}, {Bai}, {Bao}, {Bastieri}, {Bi}, {Bi},
  {Cai}, {Cai}, {Cao}, {Chang}, {Chang}, {Chen}, {Chen}, {Chen}, {Chen},
  {Chen}, {Chen}, {Chen}, {Chen}, {Chen}, {Chen}, {Chen}, {Chen}, {Chen},
  {Chen}, {Cheng}, {Cheng}, {Cui}, {Cui}, {Cui}, {D'Ettorre Piazzoli}, {Dai},
  {Dai}, {Dai}, {Danzengluobu}, {Della Volpe}, {Dong}, {Duan}, {Fan}, {Fan},
  {Fan}, {Fang}, {Fang}, {Feng}, {Feng}, {Feng}, {Feng}, {Gao}, {Gao}, {Gao},
  {Gao}, {Gao}, {Ge}, {Geng}, {Gong}, {Gou}, {Gu}, {Guo}, {Guo}, {Guo}, {Guo},
  {Guo}, {Han}, {He}, {He}, {He}, {He}, {He}, {He}, {Heller}, {Hor}, {Hou},
  {Hou}, {Hu}, {Hu}, {Hu}, {Hu}, {Huang}, {Huang}, {Huang}, {Huang}, {Huang},
  {Huang}, {Ji}, {Ji}, {Jia}, {Jiang}, {Jiang}, {Jin}, {Ke}, {Kuleshov},
  {Levochkin}, {Li}, {Li}, {Li}, {Li}, {Li}, {Li}, {Li}, {Li}, {Li}, {Li},
  {Li}, {Li}, {Li}, {Li}, {Li}, {Li}, {Li}, {Li}, {Liang}, {Liang}, {Lin},
  {Liu}, {Liu}, {Liu}, {Liu}, {Liu}, {Liu}, {Liu}, {Liu}, {Liu}, {Liu}, {Liu},
  {Liu}, {Liu}, {Liu}, {Liu}, {Liu}, {Long}, {Lu}, {Lv}, {Ma}, {Ma}, {Ma},
  {Mao}, {Masood}, {Min}, {Mitthumsiri}, {Montaruli}, {Nan}, {Pang},
  {Pattarakijwanich}, {Pei}, {Qi}, {Qi}, {Qiao}, {Qin}, {Ruffolo}, {Rulev},
  {Saiz}, {Shao}, {Shchegolev}, {Sheng}, {Shi}, {Song}, {Stenkin}, {Stepanov},
  {Su}, {Sun}, {Sun}, {Sun}, {Tam}, {Tang}, {Tian}, {Wang}, {Wang}, {Wang},
  {Wang}, {Wang}, {Wang}, {Wang}, {Wang}, {Wang}, {Wang}, {Wang}, {Wang},
  {Wang}, {Wang}, {Wang}, {Wang}, {Wang}, {Wang}, {Wang}, {Wang}, {Wang},
  {Wang}, {Wei}, {Wei}, {Wei}, {Wen}, {Wu}, {Wu}, {Wu}, {Wu}, {Wu}, {Xi},
  {Xia}, {Xia}, {Xiang}, {Xiao}, {Xiao}, {Xiao}, {Xin}, {Xin}, {Xing}, {Xu},
  {Xu}, {Xue}, {Yan}, {Yan}, {Yang}, {Yang}, {Yang}, {Yang}, {Yang}, {Yang},
  {Yang}, {Yao}, {Yao}, {Ye}, {Yin}, {Yin}, {You}, {You}, {Yu}, {Yuan}, {Zeng},
  {Zeng}, {Zeng}, {Zeng}, {Zha}, {Zhai}, {Zhang}, {Zhang}, {Zhang}, {Zhang},
  {Zhang}, {Zhang}, {Zhang}, {Zhang}, {Zhang}, {Zhang}, {Zhang}, {Zhang},
  {Zhang}, {Zhang}, {Zhang}, {Zhang}, {Zhang}, {Zhang}, {Zhang}, {Zhao},
  {Zhao}, {Zhao}, {Zhao}, {Zhao}, {Zheng}, {Zheng}, {Zhou}, {Zhou}, {Zhou},
  {Zhou}, {Zhou}, {Zhou}, {Zhu}, {Zhu}, {Zhu}, {Zhu}, \& {Zuo}}]{lhaaso21}
{Lhaaso Collaboration}, {Cao}, Z., {Aharonian}, F., {et~al.} 2021, Science,
  373, 425, \dodoi{10.1126/science.abg5137}

\bibitem[{{Lucek} \& {Bell}(2000)}]{lucek00}
{Lucek}, S.~G., \& {Bell}, A.~R. 2000, \mnras, 314, 65,
  \dodoi{10.1046/j.1365-8711.2000.03363.x}

\bibitem[{{MAGIC Collaboration} {et~al.}(2019{\natexlab{a}}){MAGIC
  Collaboration}, {Acciari}, {Ansoldi}, {Antonelli}, {Arbet Engels}, {Baack},
  {Babi{\'c}}, {Banerjee}, {Barres de Almeida}, {Barrio}, {Becerra
  Gonz{\'a}lez}, {Bednarek}, {Bellizzi}, {Bernardini}, {Berti}, {Besenrieder},
  {Bhattacharyya}, {Bigongiari}, {Biland}, {Blanch}, {Bonnoli},
  {Bo{\v{s}}njak}, {Busetto}, {Carosi}, {Carosi}, {Ceribella}, {Chai},
  {Chilingaryan}, {Cikota}, {Colak}, {Colin}, {Colombo}, {Contreras},
  {Cortina}, {Covino}, {D'Amico}, {D'Elia}, {da Vela}, {Dazzi}, {de Angelis},
  {de Lotto}, {Delfino}, {Delgado}, {Depaoli}, {di Pierro}, {di Venere}, {Do
  Souto Espi{\~n}eira}, {Dominis Prester}, {Donini}, {Dorner}, {Doro},
  {Elsaesser}, {Fallah Ramazani}, {Fattorini}, {Fern{\'a}ndez-Barral},
  {Ferrara}, {Fidalgo}, {Foffano}, {Fonseca}, {Font}, {Fruck}, {Fukami},
  {Gallozzi}, {Garc{\'\i}a L{\'o}pez}, {Garczarczyk}, {Gasparyan}, {Gaug},
  {Giglietto}, {Giordano}, {Godinovi{\'c}}, {Green}, {Guberman}, {Hadasch},
  {Hahn}, {Herrera}, {Hoang}, {Hrupec}, {H{\"u}tten}, {Inada}, {Inoue},
  {Ishio}, {Iwamura}, {Jouvin}, {Kerszberg}, {Kubo}, {Kushida}, {Lamastra},
  {Lelas}, {Leone}, {Lindfors}, {Lombardi}, {Longo}, {L{\'o}pez},
  {L{\'o}pez-Coto}, {L{\'o}pez-Oramas}, {Loporchio}, {Machado de Oliveira
  Fraga}, {Maggio}, {Majumdar}, {Makariev}, {Mallamaci}, {Maneva}, {Manganaro},
  {Mannheim}, {Maraschi}, {Mariotti}, {Mart{\'\i}nez}, {Masuda}, {Mazin},
  {Mi{\'c}anovi{\'c}}, {Miceli}, {Minev}, {Miranda}, {Mirzoyan}, {Molina},
  {Moralejo}, {Morcuende}, {Moreno}, {Moretti}, {Munar-Adrover}, {Neustroev},
  {Nigro}, {Nilsson}, {Ninci}, {Nishijima}, {Noda}, {Nogu{\'e}s}, {N{\"o}the},
  {Nozaki}, {Paiano}, {Palacio}, {Palatiello}, {Paneque}, {Paoletti},
  {Paredes}, {Pe{\~n}il}, {Peresano}, {Persic}, {Prada Moroni}, {Prandini},
  {Puljak}, {Rhode}, {Rib{\'o}}, {Rico}, {Righi}, {Rugliancich}, {Saha},
  {Sahakyan}, {Saito}, {Sakurai}, {Satalecka}, {Schmidt}, {Schweizer},
  {Sitarek}, {{\v{S}}nidari{\'c}}, {Sobczynska}, {Somero}, {Stamerra}, {Strom},
  {Strzys}, {Suda}, {Suri{\'c}}, {Takahashi}, {Tavecchio}, {Temnikov},
  {Terzi{\'c}}, {Teshima}, {Torres-Alb{\`a}}, {Tosti}, {Tsujimoto}, {Vagelli},
  {van Scherpenberg}, {Vanzo}, {Vazquez Acosta}, {Vigorito}, {Vitale}, {Vovk},
  {Will}, {Zari{\'c}}, \& {Nava}}]{magic19a}
{MAGIC Collaboration}, {Acciari}, V.~A., {Ansoldi}, S., {et~al.}
  2019{\natexlab{a}}, \nat, 575, 455, \dodoi{10.1038/s41586-019-1750-x}

\bibitem[{{MAGIC Collaboration} {et~al.}(2019{\natexlab{b}}){MAGIC
  Collaboration}, {Acciari}, {Ansoldi}, {Antonelli}, {Engels}, {Baack},
  {Babi{\'c}}, {Banerjee}, {Barres de Almeida}, {Barrio}, {Becerra
  Gonz{\'a}lez}, {Bednarek}, {Bellizzi}, {Bernardini}, {Berti}, {Besenrieder},
  {Bhattacharyya}, {Bigongiari}, {Biland}, {Blanch}, {Bonnoli},
  {Bo{\v{s}}njak}, {Busetto}, {Carosi}, {Ceribella}, {Chai}, {Chilingaryan},
  {Cikota}, {Colak}, {Colin}, {Colombo}, {Contreras}, {Cortina}, {Covino},
  {D'Elia}, {da Vela}, {Dazzi}, {de Angelis}, {de Lotto}, {Delfino}, {Delgado},
  {Depaoli}, {di Pierro}, {di Venere}, {Do Souto Espi{\~n}eira}, {Dominis
  Prester}, {Donini}, {Dorner}, {Doro}, {Elsaesser}, {Fallah Ramazani},
  {Fattorini}, {Ferrara}, {Fidalgo}, {Foffano}, {Fonseca}, {Font}, {Fruck},
  {Fukami}, {Garc{\'\i}a L{\'o}pez}, {Garczarczyk}, {Gasparyan}, {Gaug},
  {Giglietto}, {Giordano}, {Godinovi{\'c}}, {Green}, {Guberman}, {Hadasch},
  {Hahn}, {Herrera}, {Hoang}, {Hrupec}, {H{\"u}tten}, {Inada}, {Inoue},
  {Ishio}, {Iwamura}, {Jouvin}, {Kerszberg}, {Kubo}, {Kushida}, {Lamastra},
  {Lelas}, {Leone}, {Lindfors}, {Lombardi}, {Longo}, {L{\'o}pez},
  {L{\'o}pez-Coto}, {L{\'o}pez-Oramas}, {Loporchio}, {Machado de Oliveira
  Fraga}, {Maggio}, {Majumdar}, {Makariev}, {Mallamaci}, {Maneva}, {Manganaro},
  {Mannheim}, {Maraschi}, {Mariotti}, {Mart{\'\i}nez}, {Mazin},
  {Mi{\'c}anovi{\'c}}, {Miceli}, {Minev}, {Miranda}, {Mirzoyan}, {Molina},
  {Moralejo}, {Morcuende}, {Moreno}, {Moretti}, {Munar-Adrover}, {Neustroev},
  {Nigro}, {Nilsson}, {Ninci}, {Nishijima}, {Noda}, {Nogu{\'e}s}, {Nozaki},
  {Paiano}, {Palatiello}, {Paneque}, {Paoletti}, {Paredes}, {Pe{\~n}il},
  {Peresano}, {Persic}, {Moroni}, {Prandini}, {Puljak}, {Rhode}, {Rib{\'o}},
  {Rico}, {Righi}, {Rugliancich}, {Saha}, {Sahakyan}, {Saito}, {Sakurai},
  {Satalecka}, {Schmidt}, {Schweizer}, {Sitarek}, {{\v{S}}nidari{\'c}},
  {Sobczynska}, {Somero}, {Stamerra}, {Strom}, {Strzys}, {Suda}, {Suri{\'c}},
  {Takahashi}, {Tavecchio}, {Temnikov}, {Terzi{\'c}}, {Teshima},
  {Torres-Alb{\`a}}, {Tosti}, {Vagelli}, {van Scherpenberg}, {Vanzo}, {Vazquez
  Acosta}, {Vigorito}, {Vitale}, {Vovk}, {Will}, {Zari{\'c}}, {Nava}, {Veres},
  {Bhat}, {Briggs}, {Cleveland}, {Hamburg}, {Hui}, {Mailyan}, {Preece},
  {Roberts}, {von Kienlin}, {Wilson-Hodge}, {Kocevski}, {Arimoto}, {Tak},
  {Asano}, {Axelsson}, {Barbiellini}, {Bissaldi}, {Dirirsa}, {Gill}, {Granot},
  {McEnery}, {Omodei}, {Razzaque}, {Piron}, {Racusin}, {Thompson}, {Campana},
  {Bernardini}, {Kuin}, {Siegel}, {Cenko}, {O'Brien}, {Capalbi}, {Da{\i}}, {de
  Pasquale}, {Gropp}, {Klingler}, {Osborne}, {Perri}, {Starling},
  {Tagliaferri}, {Tohuvavohu}, {Ursi}, {Tavani}, {Cardillo}, {Casentini},
  {Piano}, {Evangelista}, {Verrecchia}, {Pittori}, {Lucarelli}, {Bulgarelli},
  {Parmiggiani}, {Anderson}, {Anderson}, {Bernardi}, {Bolmer},
  {Caballero-Garc{\'\i}a}, {Carrasco}, {Castell{\'o}n}, {Castro Segura},
  {Castro-Tirado}, {Cherukuri}, {Cockeram}, {D'Avanzo}, {di Dato}, {Diretse},
  {Fender}, {Fern{\'a}ndez-Garc{\'\i}a}, {Fynbo}, {Fruchter}, {Greiner},
  {Gromadzki}, {Heintz}, {Heywood}, {van der Horst}, {Hu}, {Inserra}, {Izzo},
  {Jaiswal}, {Jakobsson}, {Japelj}, {Kankare}, {Kann}, {Kouveliotou}, {Klose},
  {Levan}, {Li}, {Lotti}, {Maguire}, {Malesani}, {Manulis}, {Marongiu},
  {Martin}, {Melandri}, {Micha{\l}owski}, {Miller-Jones}, {Misra}, {Moin},
  {Mooley}, {Nasri}, {Nicholl}, {Noschese}, {Novara}, {Pandey}, {Peretti},
  {P{\'e}rez Del Pulgar}, {P{\'e}rez-Torres}, {Perley}, {Piro}, {Ragosta},
  {Resmi}, {Ricci}, {Rossi}, {S{\'a}nchez-Ram{\'\i}rez}, {Selsing}, {Schulze},
  {Smartt}, {Smith}, {Sokolov}, {Stevens}, {Tanvir}, {Th{\"o}ne}, {Tiengo},
  {Tremou}, {Troja}, {de Ugarte Postigo}, {Valeev}, {Vergani}, {Wieringa},
  {Woudt}, {Xu}, {Yaron}, \& {Young}}]{magic19b}
---. 2019{\natexlab{b}}, \nat, 575, 459, \dodoi{10.1038/s41586-019-1754-6}

\bibitem[{{Matsumoto} {et~al.}(2013){Matsumoto}, {Amano}, \&
  {Hoshino}}]{matsumoto13}
{Matsumoto}, Y., {Amano}, T., \& {Hoshino}, M. 2013, \prl, 111, 215003,
  \dodoi{10.1103/PhysRevLett.111.215003}

\bibitem[{{Matsumoto} {et~al.}(2015){Matsumoto}, {Amano}, {Kato}, \&
  {Hoshino}}]{matsumoto15}
{Matsumoto}, Y., {Amano}, T., {Kato}, T.~N., \& {Hoshino}, M. 2015, Science,
  347, 974, \dodoi{10.1126/science.1260168}

\bibitem[{{Mizuno} {et~al.}(2011){Mizuno}, {Pohl}, {Niemiec}, {Zhang},
  {Nishikawa}, \& {Hardee}}]{mizuno11}
{Mizuno}, Y., {Pohl}, M., {Niemiec}, J., {et~al.} 2011, \apj, 726, 62,
  \dodoi{10.1088/0004-637X/726/2/62}

\bibitem[{{Mizuno} {et~al.}(2014){Mizuno}, {Pohl}, {Niemiec}, {Zhang},
  {Nishikawa}, \& {Hardee}}]{mizuno14}
---. 2014, \mnras, 439, 3490, \dodoi{10.1093/mnras/stu196}

\bibitem[{{Murase} {et~al.}(2012){Murase}, {Dermer}, {Takami}, \&
  {Migliori}}]{murase12}
{Murase}, K., {Dermer}, C.~D., {Takami}, H., \& {Migliori}, G. 2012, \apj, 749,
  63, \dodoi{10.1088/0004-637X/749/1/63}

\bibitem[{{Niemiec} {et~al.}(2006){Niemiec}, {Ostrowski}, \&
  {Pohl}}]{niemiec06}
{Niemiec}, J., {Ostrowski}, M., \& {Pohl}, M. 2006, \apj, 650, 1020,
  \dodoi{10.1086/506901}

\bibitem[{{Niemiec} {et~al.}(2012){Niemiec}, {Pohl}, {Bret}, \&
  {Wieland}}]{niemiec12}
{Niemiec}, J., {Pohl}, M., {Bret}, A., \& {Wieland}, V. 2012, \apj, 759, 73,
  \dodoi{10.1088/0004-637X/759/1/73}

\bibitem[{{Nishihara} {et~al.}(2010){Nishihara}, {Wouchuk}, {Matsuoka},
  {Ishizaki}, \& {Zhakhovsky}}]{nishihara10}
{Nishihara}, K., {Wouchuk}, J.~G., {Matsuoka}, C., {Ishizaki}, R., \&
  {Zhakhovsky}, V.~V. 2010, Philosophical Transactions of the Royal Society of
  London Series A, 368, 1769, \dodoi{10.1098/rsta.2009.0252}

\bibitem[{{Ohira}(2013)}]{ohira13}
{Ohira}, Y. 2013, \apjl, 767, L16, \dodoi{10.1088/2041-8205/767/1/L16}

\bibitem[{{Ohira}(2016{\natexlab{a}})}]{ohira16a}
---. 2016{\natexlab{a}}, \apj, 817, 137, \dodoi{10.3847/0004-637X/817/2/137}

\bibitem[{{Ohira}(2016{\natexlab{b}})}]{ohira16b}
---. 2016{\natexlab{b}}, \apj, 827, 36, \dodoi{10.3847/0004-637X/827/1/36}

\bibitem[{{Ohira} {et~al.}(2009){Ohira}, {Terasawa}, \& {Takahara}}]{ohira09}
{Ohira}, Y., {Terasawa}, T., \& {Takahara}, F. 2009, \apjl, 703, L59,
  \dodoi{10.1088/0004-637X/703/1/L59}

\bibitem[{{Perkins} {et~al.}(2017){Perkins}, {Ho}, {Logan}, {Zimmerman},
  {Rhodes}, {Strozzi}, {Blackfield}, \& {Hawkins}}]{perkins17}
{Perkins}, L.~J., {Ho}, D.~D.~M., {Logan}, B.~G., {et~al.} 2017, Physics of
  Plasmas, 24, 062708, \dodoi{10.1063/1.4985150}

\bibitem[{{Pohl} {et~al.}(2015){Pohl}, {Wilhelm}, \& {Telezhinsky}}]{pohl15}
{Pohl}, M., {Wilhelm}, A., \& {Telezhinsky}, I. 2015, \aap, 574, A43,
  \dodoi{10.1051/0004-6361/201425027}

\bibitem[{{Sano} {et~al.}(2013){Sano}, {Inoue}, \& {Nishihara}}]{sano13}
{Sano}, T., {Inoue}, T., \& {Nishihara}, K. 2013, \prl, 111, 205001,
  \dodoi{10.1103/PhysRevLett.111.205001}

\bibitem[{{Sano} {et~al.}(2021){Sano}, {Tamatani}, {Matsuo}, {Law}, {Morita},
  {Egashira}, {Ota}, {Kumar}, {Shimogawara}, {Hara}, {Lee}, {Sakata}, {Rigon},
  {Michel}, {Mabey}, {Albertazzi}, {Koenig}, {Casner}, {Shigemori}, {Fujioka},
  {Murakami}, \& {Sakawa}}]{sano21}
{Sano}, T., {Tamatani}, S., {Matsuo}, K., {et~al.} 2021, \pre, 104, 035206,
  \dodoi{10.1103/PhysRevE.104.035206}

\bibitem[{{Santana} {et~al.}(2014){Santana}, {Barniol Duran}, \&
  {Kumar}}]{santana14}
{Santana}, R., {Barniol Duran}, R., \& {Kumar}, P. 2014, \apj, 785, 29,
  \dodoi{10.1088/0004-637X/785/1/29}

\bibitem[{{Shin} {et~al.}(2008){Shin}, {Stone}, \& {Snyder}}]{shin08}
{Shin}, M.-S., {Stone}, J.~M., \& {Snyder}, G.~F. 2008, \apj, 680, 336,
  \dodoi{10.1086/587775}

\bibitem[{{Sironi} \& {Goodman}(2007)}]{sironi07}
{Sironi}, L., \& {Goodman}, J. 2007, \apj, 671, 1858, \dodoi{10.1086/523636}

\bibitem[{{Sironi} \& {Spitkovsky}(2012)}]{sironi12}
{Sironi}, L., \& {Spitkovsky}, A. 2012, Computational Science and Discovery, 5,
  014014, \dodoi{10.1088/1749-4699/5/1/014014}

\bibitem[{{Sironi} {et~al.}(2013){Sironi}, {Spitkovsky}, \& {Arons}}]{sironi13}
{Sironi}, L., {Spitkovsky}, A., \& {Arons}, J. 2013, \apj, 771, 54,
  \dodoi{10.1088/0004-637X/771/1/54}

\bibitem[{{Spitkovsky}(2008{\natexlab{a}})}]{spitkovsky08a}
{Spitkovsky}, A. 2008{\natexlab{a}}, \apjl, 682, L5, \dodoi{10.1086/590248}

\bibitem[{{Spitkovsky}(2008{\natexlab{b}})}]{spitkovsky08b}
---. 2008{\natexlab{b}}, \apjl, 673, L39, \dodoi{10.1086/527374}

\bibitem[{{Stone} \& {Norman}(1992)}]{stone92}
{Stone}, J.~M., \& {Norman}, M.~L. 1992, \apjl, 390, L17,
  \dodoi{10.1086/186361}

\bibitem[{{Stone} {et~al.}(2020){Stone}, {Tomida}, {White}, \&
  {Felker}}]{stone20}
{Stone}, J.~M., {Tomida}, K., {White}, C.~J., \& {Felker}, K.~G. 2020, \apjs,
  249, 4, \dodoi{10.3847/1538-4365/ab929b}

\bibitem[{{Tanaka} \& {Takahara}(2011)}]{tanaka11}
{Tanaka}, S.~J., \& {Takahara}, F. 2011, \apj, 741, 40,
  \dodoi{10.1088/0004-637X/741/1/40}

\bibitem[{{Tomita} \& {Ohira}(2016)}]{tomita16}
{Tomita}, S., \& {Ohira}, Y. 2016, \apj, 825, 103,
  \dodoi{10.3847/0004-637X/825/2/103}

\bibitem[{{Tomita} {et~al.}(2019){Tomita}, {Ohira}, \& {Yamazaki}}]{tomita19}
{Tomita}, S., {Ohira}, Y., \& {Yamazaki}, R. 2019, \apj, 886, 54,
  \dodoi{10.3847/1538-4357/ab4a10}

\bibitem[{{Uchiyama} {et~al.}(2007){Uchiyama}, {Aharonian}, {Tanaka},
  {Takahashi}, \& {Maeda}}]{uchiyama07}
{Uchiyama}, Y., {Aharonian}, F.~A., {Tanaka}, T., {Takahashi}, T., \& {Maeda},
  Y. 2007, \nat, 449, 576, \dodoi{10.1038/nature06210}

\bibitem[{{Weibel}(1959)}]{weibel59}
{Weibel}, E.~S. 1959, \prl, 2, 83, \dodoi{10.1103/PhysRevLett.2.83}

\bibitem[{{White} {et~al.}(2016){White}, {Stone}, \& {Gammie}}]{white16}
{White}, C.~J., {Stone}, J.~M., \& {Gammie}, C.~F. 2016, \apjs, 225, 22,
  \dodoi{10.3847/0067-0049/225/2/22}

\bibitem[{{Yokoyama} \& {Ohira}(2020)}]{yokoyama20}
{Yokoyama}, S.~L., \& {Ohira}, Y. 2020, \apj, 897, 50,
  \dodoi{10.3847/1538-4357/ab93c3}

\bibitem[{{Zhang} {et~al.}(2018){Zhang}, {Murase}, {Kimura}, {Horiuchi}, \&
  {M{\'e}sz{\'a}ros}}]{zhang18}
{Zhang}, B.~T., {Murase}, K., {Kimura}, S.~S., {Horiuchi}, S., \&
  {M{\'e}sz{\'a}ros}, P. 2018, \prd, 97, 083010,
  \dodoi{10.1103/PhysRevD.97.083010}

\bibitem[{{Zhdankin} {et~al.}(2018){Zhdankin}, {Uzdensky}, {Werner}, \&
  {Begelman}}]{zhdankin18}
{Zhdankin}, V., {Uzdensky}, D.~A., {Werner}, G.~R., \& {Begelman}, M.~C. 2018,
  \apjl, 867, L18, \dodoi{10.3847/2041-8213/aae88c}

\bibitem[{{Zhou} {et~al.}(2021){Zhou}, {Williams}, {Ramaprabhu}, {Groom},
  {Thornber}, {Hillier}, {Mostert}, {Rollin}, {Balachandar}, {Powell},
  {Mahalov}, \& {Attal}}]{zhou21}
{Zhou}, Y., {Williams}, R. J.~R., {Ramaprabhu}, P., {et~al.} 2021, Physica D
  Nonlinear Phenomena, 423, 132838, \dodoi{10.1016/j.physd.2020.132838}

\end{thebibliography}
\bibliographystyle{aasjournal}



\end{document}